\documentclass[12pt]{article}
\usepackage[totalwidth=460truept,totalheight=600truept]{geometry}
\usepackage{amsfonts,latexsym,amssymb,amsmath,graphicx}
\usepackage[hidelinks]{hyperref}

\global\arraycolsep=1truept

\begin{document}

\bigskip \phantom{C}

\vskip 2truecm

\begin{center}
{\huge \textbf{On The Nature}}

\vskip.4truecm

{\huge \textbf{Of The Higgs Boson}}

\vskip1truecm

\textsl{Damiano Anselmi}

\vskip .1truecm

\textit{Dipartimento di Fisica ``Enrico Fermi'', Universit\`{a} di Pisa, }

\textit{Largo B. Pontecorvo 3, 56127 Pisa, Italy}

\textit{and INFN, Sezione di Pisa,}

\textit{Largo B. Pontecorvo 3, 56127 Pisa, Italy}

damiano.anselmi@unipi.it

\vskip1truecm

\textbf{Abstract}
\end{center}

Several particles are not observed directly, but only through their decay
products. We consider the possibility that they might be fakeons, i.e. fake
particles, which mediate interactions but are not asymptotic states. A
crucial role to determine the true nature of a particle is played by the
imaginary parts of the one-loop radiative corrections, which are affected in
nontrivial ways by the presence of fakeons in the loop. The knowledge we
have today is sufficient to prove that most non directly observed particles
are true physical particles. However, in the case of the Higgs boson the
possibility that it might be a fakeon remains open. The issue can be
resolved by means of precision measurements in existing and future
accelerators.

\vfill\eject

The Higgs boson has unique features. For example, it is a scalar field,
unlike every other field of the standard model. Its key role is to trigger a
crucial mechanism that gives masses to the particles. While it solves many
problems, it leaves other questions unanswered. In this paper we study the
possibility that it might hide a little secret. Specifically, the Higgs
boson might be a \textquotedblleft fake particle\textquotedblright , i.e. an
entity that resembles a true particle in various physical processes, but
cannot be observed directly.

In quantum field theory, the poles of the free propagators are usually
quantized by means of the Feynman prescription \cite{peskin}. In that case,
they describe physical particles. An alternative quantization prescription
is able to quantize them as fake particles \cite{LWgrav}, or
\textquotedblleft fakeons\textquotedblright\ \cite{fakeons}. The fakeons are
important in quantum gravity, because they allow us to build a consistent
theory that is both unitary and renormalizable \cite{LWgrav} (see also \cite%
{UVQG,absograv}).

A fakeon simulates a physical particle when it mediates interactions or
decays into physical particles. However, it is not an asymptotic state,
because unitarity requires to project the fakeons away from the physical
spectrum. In other words, a fakeon cannot be detected directly. An important
physical prediction due to the fakeons is the violation of microcausality,
which occurs at energies larger than their masses.

Quantum gravity predicts that at least one fakeon exists in nature \cite%
{LWgrav}. It has spin 2, it is described by a symmetric tensor $\chi _{\mu
\nu }$ and its mass $m_{\chi }$ could be much smaller than the Planck mass.
Its free propagator has a negative residue at the pole, so $\chi _{\mu \nu }$
is a \textquotedblleft fakeon minus\textquotedblright\ \cite{causalityQG}
and its dynamically generated width $\Gamma _{\chi }$ is negative.

The spin-2 gravifakeon is necessary to make the quantization of gravity
consistent. In other sectors of high-energy physics, like the standard model
in flat space, as well as its extensions, there might be no need of fake
particles. However, if one fakeon exists in nature, it might not be the only
one. Are there any other fakeons, maybe in the realm of the standard model?
In this paper we provide enough arguments to exclude this possibility for
most particles, but the cases of the Higgs boson and a few other particles
remain unresolved.

Standard model extensions can be built by adding physical and fake
particles, as long as they satisfy the conditions for the cancellation of
the gauge anomalies and their masses are large enough to avoid conflict with
the data. We do not explore these possibilities here, although they might
have interesting applications. Instead, we inquire whether the particles
that have already been identified so far are physical or fake.

Some particles, like the photon and the electron, are observed directly, so
they are physical. Several other particles have not been observed directly,
and probably will not be for a long time. We can mention the intermediate
bosons, the Higgs boson, the quarks, the gluons and the neutrinos. All of
these are potentially fakeons. Since the definition of direct observation of
a particle is to some extent debatable, we prefer to determine the true
nature of all particles, including the photon and the electron, by means of
indirect, more objective methods.

We show that, as of today, we have enough data to ensure that most particles
are physical. However, we are unable to settle the matter in the cases of
the Higgs boson,\ the top quark, the gluons and the right neutrinos. With
the exception of the right neutrinos, the missing answers can be provided by
precision measurements to be made in existing and/or future accelerators.

Before proceeding, let us recall a few properties. Physical and fake
particles are quantized by means of different prescriptions, which are the
Feynman prescription and the fakeon prescription. Introducing infinitesimal
widths $\epsilon $ and $\mathcal{E}$, the unprescribed propagator $%
1/(p^{2}-m^{2})$ is turned into 
\begin{equation}
\frac{1}{p^{2}-m^{2}+i\epsilon },\qquad \frac{p^{2}-m^{2}}{(p^{2}-m^{2})^{2}+%
\mathcal{E}^{4}},  \label{prescr}
\end{equation}%
respectively. Note that, by convention, $\epsilon $ and $\mathcal{E}$ have
different dimensions. The fakeon propagator vanishes on shell (which means
for $p^{2}=m^{2}$) for every $\mathcal{E}>0$. This is a sign that is does
not propagate a physical particle. Instead, the Feynman propagator blows up
on shell. Off-shell, for \ $|p^{2}-m^{2}|\gg \epsilon ,\mathcal{E}^{2}$, the
two prescriptions are equivalent.

The sign in front of the Feynman propagator must be positive, otherwise it
propagates a ghost, instead of a physical particle. Instead, the sign in
front of the fakeon propagator can be either positive or negative, which
distinguishes the \textquotedblleft fakeon plus\textquotedblright\ from the
\textquotedblleft fakeon minus\textquotedblright\ \cite{causalityQG}. The
dynamically generated width $\Gamma $ of a fakeon plus (minus) is positive
(negative).

In the right expression of (\ref{prescr}) the poles of $1/(p^{2}-m^{2})$ are
split into pairs of complex conjugates poles. Inside the Feynman diagrams, the
loop energy $p^{0}$ must be integrated along a path that passes under the
left pair and over the right pair. The fakeon prescription needs to be
specified by a number of other instructions to compute the loop integrals, which we do not review here and
have been recently summarized in ref. \cite{causalityQG}. We mention, in particular,
the average continuation, which is a nonanalytic operation to circumvent the thresholds that involve fakeons.
In the end, the
most important property is that the fake particles can be consistently
projected away from the physical spectrum.

At the tree level the difference between the two options (physical or fake)
is just the projection, which has no impact on the particles that are not
detected directly, like the vector bosons and the Higgs boson. The true
nature of these particles can be established by analyzing the radiative
corrections to the scattering processes. Since the physical spectrum is
defined with respect to the broken phase, $SU(2)$ invariance is not very
helpful. This means that the quantization prescriptions of the $W^{\pm }$
bosons, the $Z$ boson, the photon $\gamma $ and the Higgs field $\eta $ are
in principle unrelated to one another.

\begin{figure}[t]
\begin{center}
\includegraphics[width=5truecm]{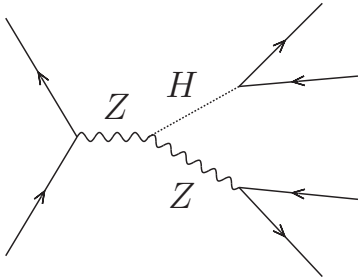}
\end{center}
\caption{Higgs decay processes}
\label{HiggsDecay}
\end{figure}

To begin with, consider a process like the one shown in fig. \ref{HiggsDecay}%
. Since $H$ and $Z$ decay, both quantization prescriptions (\ref{prescr})
give the same prediction. This is a typical case where $H$ and $Z$ simulate
real particles even if they are quantized as fakeons.

Specifically, it is possible to show \cite{absograv,causalityQG} that if we
take a fakeon $F$ and resum the powers of its dynamical width $\Gamma _{F}$
into the so-called dressed propagator, the imaginary part of (minus) the $F$
dressed propagator tends to 
\begin{equation}
\pi Z_{F}\delta (p^{2}-m_{F}^{2})  \label{dress}
\end{equation}%
in the limit $|\Gamma _{F}|\rightarrow 0$, as if the fakeon $F$ \textit{were
indeed a physical particle} of mass $m_{F}$, where $Z_{F}$ is the
normalization factor. Applying these arguments to the final states of the
process of fig. \ref{HiggsDecay}, we see no difference between true and fake 
$H$ and $Z$ particles.

\begin{figure}[t]
\begin{center}
\includegraphics[width=10truecm]{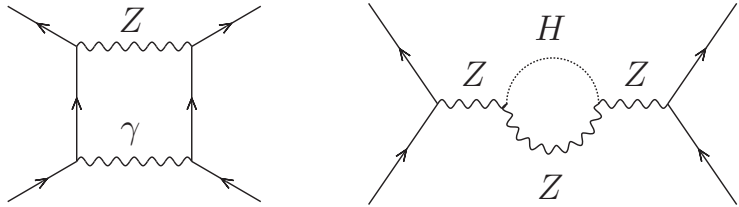}
\end{center}
\caption{Relevant processes with $Z$ and $H$ bosons in loops}
\label{HiggsDiag}
\end{figure}

At one loop, the two prescriptions give significantly different
results for the imaginary parts $\mathcal{I}$ of the radiative corrections
to the transition amplitudes above suitable thresholds \cite{LWgrav,fakeons}, while the real parts
coincide in the two cases. The relevant diagrams are
self-energies, triangle diagrams and box diagrams. See fig. \ref{HiggsDiag}
for examples and refs. \cite{hollik} for explicit formulas. We need to
pay attention to the diagrams that contain at least one virtual fakeon and
estimate the orders of magnitude of the various types of contributions. An
important point is that the processes we are considering are far from the
resonance peaks. In such conditions, the contributions of the three types of
diagrams are of the same orders, so in many cases we can concentrate on the
self-energies with no loss of generality.

Consider a self-energy diagram $\mathcal{B}$ with internal legs of masses $%
m_{1}$ and $m_{2}$. The imaginary part $\mathcal{I}$ of $-i\mathcal{B}$ is
equal to zero if an internal leg is quantized as a fakeon \cite%
{LWgrav,fakeons,UVQG,absograv,causalityQG}, while it is proportional to%
\begin{equation}
\theta \left( s-(m_{1}+m_{2})^{2}\right) \sqrt{1-\frac{(m_{1}+m_{2})^{2}}{s}}%
\sqrt{1-\frac{(m_{1}-m_{2})^{2}}{s}}  \label{ima}
\end{equation}%
if both internal legs are quantized as physical particles, where $s$ is the
center-of-mass energy squared. The real part of $-i\mathcal{B}$ is the same
with both quantization prescriptions. Typically, when neither of the
particles circulating in the loop are fakeons, $\mathcal{I}$ and the real
part of $-i\mathcal{B}$ are of the same order, when $s$ is larger than the
physical threshold $(m_{1}+m_{2})^{2}$.

Let $\mathcal{I}_{cd}^{ab}$ denote the imaginary part of $-i$ times the
bubble diagram that has $a$, $b$ as external legs and $c$, $d$ as
circulating particles. If $a=b$, we just write $\mathcal{I}_{cd}^{a}$.
Consider $\mathcal{I}_{W^{+}W^{-}}^{\gamma }$, $\mathcal{I}_{W^{+}W^{-}}^{Z}$%
, $\mathcal{I}_{W^{+}W^{-}}^{\gamma Z}$ and $\mathcal{I}_{ZH}^{Z}$. Since
they have different thresholds, or depend on $s$ in different ways, it is
possible to analyze their contributions separately in precision
measurements. The imaginary parts $\mathcal{I}_{W^{+}W^{-}}^{\gamma }$, $%
\mathcal{I}_{W^{+}W^{-}}^{Z}$ and $\mathcal{I}_{W^{+}W^{-}}^{\gamma Z}$
contribute to the cross section $\sigma (e^{+}e^{-}\rightarrow $ leptons,
hadrons$)$ and their thresholds $2m_{W}$ are in the range of energies
spanned for example by LEP\ II. Since no unexpected behavior has been
noted (and the data of LEP\ II are precise enough), we infer that $%
\mathcal{I}_{W^{+}W^{-}}^{ab}$ are nonvanishing, hence the $W$ bosons are
physical and not fake. The same can be said of the left neutrinos $\nu $,
from $\mathcal{I}_{\nu \bar{\nu}}^{Z}$, the charged leptons $\ell $, from $%
\mathcal{I}_{\ell \bar{\ell}}^{\gamma }$, $\mathcal{I}_{\ell \bar{\ell}}^{Z}$%
, $\mathcal{I}_{\ell \bar{\ell}}^{\gamma Z}$, and all the quarks $q$ but the
top one, from $\mathcal{I}_{q\bar{q}}^{\gamma }$, $\mathcal{I}_{q\bar{q}%
}^{Z} $, $\mathcal{I}_{q\bar{q}}^{\gamma Z}$.

The case of the top quark $t$ remains unresolved. The self-energy
contributions $\mathcal{I}_{t\bar{t}}^{\gamma }$, $\mathcal{I}_{t\bar{t}%
}^{Z} $ and $\mathcal{I}_{t\bar{t}}^{\gamma Z}$ with circulating tops have
been missed by LEP\ II, due to their thresholds $2m_{t}\sim 346$GeV.
However, they can be studied in precision measurements with some effort of
data analysis and background subtraction at LHC \cite{atlas} and HiLumi \cite%
{HiLumi}, or, more directly, at the International Linear Collider \cite{ILC}%
, the Compact Linear Collider \cite{CLIC}, the Future Circular Collider \cite%
{FCC} and the Circular Electron Positron Collider \cite{CEPC}, if they will
be eventually built \cite{Mangano}.

The analysis just made, based on the self-energies, leaves out the photon $%
\gamma $ and the $Z$ boson. We can prove that they are not fake by
considering the box diagram of fig. \ref{HiggsDiag}. Here the imaginary part
associated with the vertical cut has a threshold equal to $m_{Z}$, so it
contributes to the processes studied at LEP II. Since no discrepancies with
respect to the predictions of the usual quantization prescription have been
reported, we infer that both $\gamma $ and $Z$ are physical. As far as the $%
Z $ boson is concerned, we can also consider the box diagram with $\gamma $
replaced by a second $Z$, since the threshold $2m_{Z}$ has also been
exceeded by LEP\ II.

Let us now consider the Higgs boson. If the Higgs field were a fakeon, it
would have to be a fakeon plus, since the possibility that it might be a
fakeon minus is excluded. Indeed, to turn it into a fakeon minus, we would
have to flip the sign of its kinetic term $(D_{\mu }H^{\dag })(D^{\mu }H)$.
Then the squared masses of $Z$ and $W^{\pm }$ would also turn into their
opposites. However, it is not possible to quantize takyons as fakeons [both
prescriptions (\ref{prescr}) must have $m^{2}>0$], so we would have to flip
the signs of the of $Z$ and $W^{\pm }$ kinetic terms as well. That would
force us to quantize $Z$ and $W^{\pm }$ as fakeons (since the Feynman
prescription would turn them into ghosts), which is contrary to the results
obtained above. In the end, we remain with just two possibilities: the Higgs
boson is a physical particle or a fakeon plus.

To decide which it is, consider the cross sections $\sigma
(e^{+}e^{-}\rightarrow \gamma \ell j)$, where $\gamma \ell j$ denotes any
final state made of photons, leptons and/or jets. If $H$ is physical, the
imaginary part $\mathcal{I}_{ZH}^{Z}$ of the $Z$ self-energy with a Higgs
field (see the right diagram of fig. \ref{HiggsDiag}) starts contributing
from $\sqrt{s}\gtrsim m_{Z}+m_{H}=216$GeV. Enough above the threshold (say,
at $\sqrt{s}\sim 240$GeV) the contribution of $\mathcal{I}_{ZH}^{Z}$ is
comparable to the one of the real part. It is also comparable to the
contributions of the imaginary parts of the other main $Z$ self-energy
diagrams, like $\mathcal{I}_{W^{+}W^{-}}^{Z}$.

Thus, the difference between a physical Higgs boson and a fake one is
important enough to be noted, whenever a self-energy diagram like the one
of fig. \ref{HiggsDiag} contributes and the experiment is sensitive to it.
If $\mathcal{I}_{ZH}^{Z}$ is found to be nonvanishing, then both $Z$ and $H$
are not fake. Instead, if $\mathcal{I}_{ZH}^{Z}$ is found to vanish, we
conclude that $H$ must be fake, since we have already proved that $Z$ is not
fake.

If LEP II had not stopped right below the threshold $m_{Z}+m_{H}$, we would
already know the answer to this question. At present, the only possibility
to fill this gap is to perform precision measurements\ at LHC or wait for
HiLumi, ILC, CLIC, FCC or CEPC.

\begin{figure}[t]
\begin{center}
\includegraphics[width=6truecm]{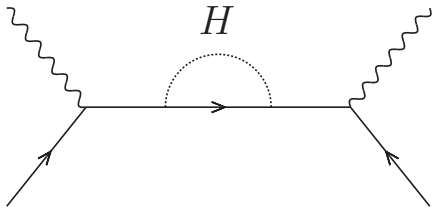}
\end{center}
\caption{Compton-like process for fermion self-energy with Higgs boson}
\label{compton}
\end{figure}

Other potentially relevant diagrams are the fermion self-energies that
involve a virtual Higgs boson. The fermions $f$ must also be virtual, to
turn on the imaginary part $\mathcal{I}_{fH}^{f}$, whose threshold is $%
m_{f}+m_{H}$. These self-energy diagrams contribute for example to the
Compton-like process of fig. \ref{compton}. However, the couplings of the
fermions to the Higgs boson are suppressed by a factor $m_{f}/v$, where $v$
is the Higgs vev, and this ratio is squared in the diagram. The resulting
contribution $\mathcal{I}_{fH}^{f}$ is too small to be observed in all cases
apart from the one of the top quark, where the threshold raises to about $%
300 $GeV. The virtual top quark can be produced by top-gluon and top-photon
interactions, as well as from the pairs $W^{+}b$ and $Zt$. These are processes
that can be studied at LHC.

Replacing $H$ with a gluon in fig. \ref{compton}, we obtain a contribution
that allows us to test whether the gluons are physical or fake. In that
case, it is enough to reach energies that are a bit larger than the quark
mass and we can use any type of quark we want. We are not aware of data that
can be immediately analyzed to obtain an answer in this case, but it is
another problem that can be studied at LHC. 

If we replace $Z$ or $\gamma $, or both, with $H$ in the box diagram of fig. %
\ref{HiggsDiag} we obtain another interesting diagram with a virtual Higgs
boson circulating in a loop. Then the fermions can only be top quarks,
since, for the reasons recalled before, the couplings of $H$ to the other
fermions are too small.

In conclusion, theoretical arguments and experimental evidence ensure that
no particles of the standard model are fakeons, apart from possibly the
Higgs boson and the top quark, the gluons and the right neutrinos. The top
quark is related to other quarks by (approximate) family symmetries, which
may suggest that it is probably physical. On the other hand, the possibility
that the Higgs boson is a fakeon is more intriguing, given the peculiarities
of this particle. All cases, apart from the one of the right neutrinos, can
be settled in future experiments or by performing precision measurements at
LHC. If one or more particles of the standard model turn out to be fakeons,
it becomes interesting to devise specific experiments to search for the
first signs of violations of microcausality.

\vskip12truept \noindent {\large \textbf{Acknowledgments}} \nopagebreak\vskip%
2truept \nopagebreak We are grateful to U. Aglietti, M. Grazzini, M. Piva
and A. Strumia for useful discussions.

\end{document}